# Toward High-Throughput Artificial Intelligence-Based Segmentation in Oncological PET Imaging

Fereshteh Yousefirizi[1*], Abhinav K. Jha[2,3], Julia Brosch-Lenz[1], Babak Saboury[4,5,6], Arman Rahmim[7,8,9]

[1]Department of Integrative Oncology, BC Cancer Research Institute, Vancouver, BC, Canada
[2]Department of Biomedical Engineering, Washington University in St. Louis, MO, USA
[3]Mallinckrodt Institute of Radiology, Washington University in St. Louis
[4]Department of Radiology and Imaging Sciences, Clinical Center, National Institutes of Health, Bethesda, MD, USA
[5]Department of Computer Science and Electrical Engineering, University of Maryland Baltimore County, Baltimore, MD, USA,
[6]Department of Radiology, Hospital of the University of Pennsylvania, Philadelphia, PA, USA
[7]Departments of Radiology and Physics, University of British Columbia
[8]Senior Scientist & Provincial Medical Imaging Physicist, BC Cancer
[9]BC Cancer Research Institute

Corresponding author: frizi@bccrc.ca

**Synopsis:** Artificial intelligence (AI) techniques for image-based segmentation have garnered much attention in recent years. Convolutional neural networks (CNNs) have shown impressive results and potential towards fully automated segmentation in medical imaging, and particularly PET imaging. To cope with the limited access to annotated data needed in supervised AI methods, given tedious and prone-to-error manual delineations, semi-supervised and unsupervised AI techniques have also been explored for segmentation of tumors or normal organs in single and bi-modality scans. This work provides a review of existing AI techniques for segmentation tasks and the evaluation criteria for translational AI-based segmentation efforts towards routine adoption in clinical workflows.

**Key Points:**
- The need for an automatic segmentation technique to support oncologic diagnosis as well as to assess the progression free survival analysis by radiomics is vital.
- The lack of annotated data and publicly available data affect the generalizability of AI techniques that have been developed to this aim.
- Semi-supervised or unsupervised AI techniques can be used to tackle the data scarcity and have potential to improve consistency and quality of annotated data

**Disclosure Statement:** Authors do not have anything to disclose regarding conflict of interest with respect to this manuscript.

## 1.   Introduction

An array of artificial intelligence (AI) techniques in the field of medical imaging have emerged in the past decade for automated image segmentation [1]. Medical image segmentation seeks to extract regions with specific anatomical and/or functional features and to classify the pixels (voxels) in terms of grey level, spatial or textural features [2] [3]. The pixels (voxels) may be segmented with varied amounts of uncertainty, given contextual information.

Accurate segmentation is also crucial for external beam therapy planning. In the last decade, the valuable role of radiomics [4] [5] for image assessment and outcome prediction has been reported, for which segmentation is a vital step [6-9]. In clinical workflows, in the context of radiopharmaceutical therapies, segmentation of PET and/or SPECT images is also needed for image-based dosimetry as well as quantification of therapy response based on pre- and post-therapeutic images. To streamline the tedious, prone-to-error and subjective task of manual delineations (e.g.

(leading to inter- and intra-observer variabilities), there have been significant efforts towards automated tumor segmentation [10-13].

Considering the quality of annotations, weak supervision can be categorized as follows: (1) incomplete supervision: when limited annotated data are provided in the training set, (2) inexact supervision: when bounding boxes and image-level annotations are provided, (3) inaccurate supervision: where the provided labels are not always ground-truth [14].

In this review, we consider supervised, weakly supervised (as generalization of semi-supervised techniques) and un-supervised AI techniques that have been used for tumor or normal organ segmentation in oncological PET and PET/CT imaging (more details in section 2). Translating AI techniques into routinely employed clinical workflows requires collaboration between AI researchers, clinicians, and predefined frameworks for evaluating these techniques to be integrated into clinical applications [15]. We outline the needed steps for AI techniques to be applicable in clinical workflows in section 3. We conclude this paper by a series of considerations for an AI technique to be applicable in the clinical workflow and future directions for automated segmentation.

## 2. AI Techniques for Image Segmentation in oncological PET imaging

Metabolic tumor volume (MTV) refers to the volume of the segmented tumor in FDG PET images. It has also been referred to as metabolically active tumor volume (MATV) [16]. There are significant studies on PET imaging using other tracers; e.g. PSMA PET, in which case this is referred to as molecular tumor volume (MTV). MTV is an important metric for response assessment and outcome prediction [17]. It has also been referred to as metabolically active tumor volume (MTV), and total MTV (TMTV) if the metastatic regions and lymph nodes are taken into consideration. Most existing studies report techniques for primary tumor segmentation, while for TMTV, accurate segmentation of metastatic regions and/or lymph nodes is also needed. As an example, TMTV is a significant prognostic factor in a range of lymphomas (diffuse large B-cell lymphoma (DLBCL), primary mediastinal B-cell lymphoma (PMBCL) and Hodgkin's lymphoma). Due to the small size of metastatic regions, their variant locations and different tumor-to-background ratios, segmentation of metastatic regions and lymph nodes is a challenging task. Figure 1 depicts an AI application for the quantification of whole-body tumor volume: AI based segmentation and differentiation between tumor lesions and physiological tracer uptake in FDG PET and PSMA PET images are shown. Segmentation techniques range from 2D to volumetric segmentations to assess the entire tumor (bulk) and/or normal organs. Different levels of supervision can be used for training a segmentation model from pixel/voxel-level annotations in supervised learning, and image-level or inaccurate annotations in weakly-supervised learning, to no annotations in unsupervised learning (Figure 2) [18].

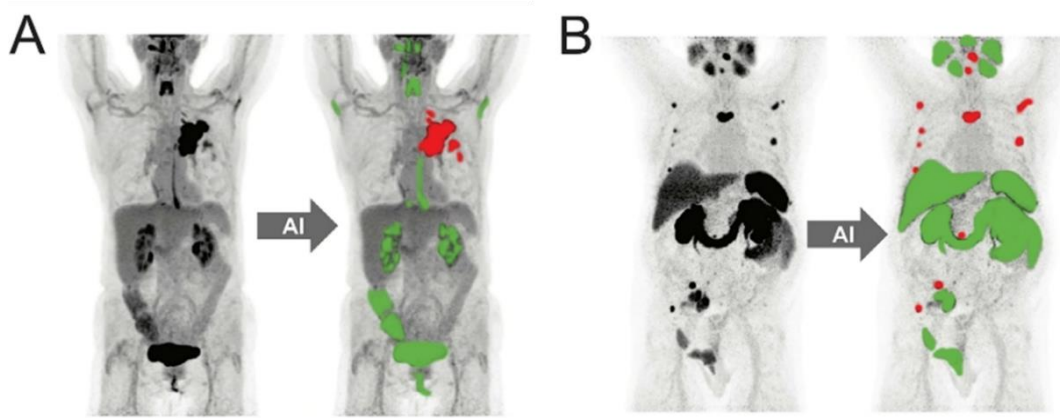

Figure 1: AI framework assists nuclear medicine experts in the reading of whole-body scans. Example segmentations of tumor lesions and physiological tracer uptake for FDG PET imaging of lung cancer (A) or PSMA PET imaging of prostate cancer (B). Physiological uptake is shown in green, while pathological uptake is in red. (Re-used with permission [19])

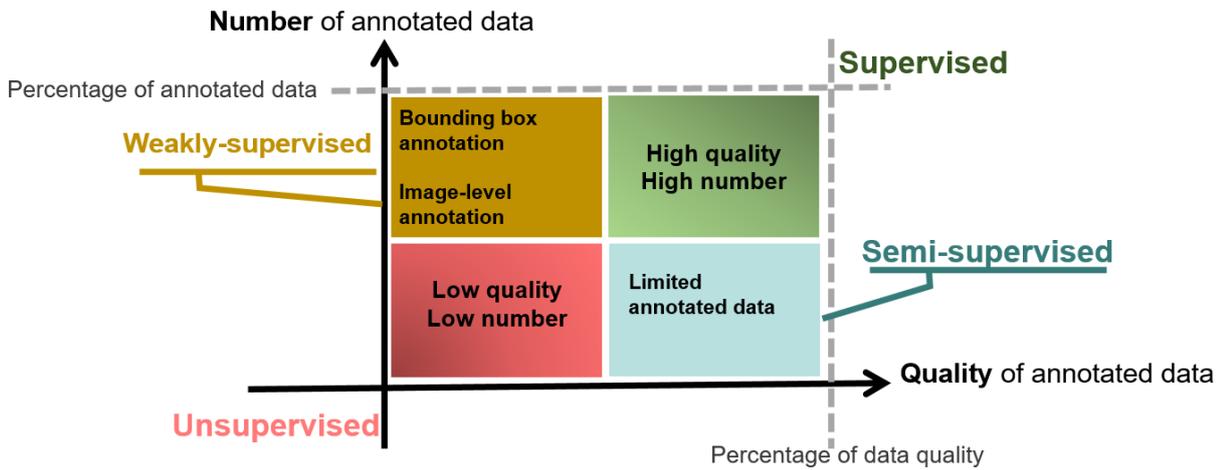

Figure 2: Different levels of supervision for training a segmentation model. We emphasize that it is possible to consider weakly supervised techniques to encompass data that are limited in quality (along x-axis); and semi-supervised techniques to encompass data that are limited in number (along y-axis) i.e. weakly supervised techniques can be thought as generalization of semi-supervised techniques. With high quality and high number of annotated data within the training data, one moves towards fully supervised AI techniques.

Supervised segmentation techniques are applicable if pixel-level annotations are available. In the case of limited annotated data (i.e. limited in number), semi-supervised techniques are helpful. If only bounding-box or image-level weak annotations (e.g. annotations of objects and attributes without spatial localization or associations between them) are available, weakly-supervised techniques can be applicable. When no labels are available, un-supervised techniques would be the solution. Figure 2 shows the different learning techniques for different levels of supervision. We note that it is possible to consider weakly-supervised techniques to encompass data that are limited in quality (along x-axis); and semi-supervised techniques for data that are limited in number (y-axis) see [14]). As such, weakly-supervised techniques can be thought as generalization of semi-supervised techniques. Overall, it is worth noting that since large numbers of un- or weakly-annotated images can be available, they can be potentially combined/cascaded with small yet well-annotated images [20].

Deep learning (DL) techniques, especially CNNs have shown to be effective for medical image segmentation [2] [21], specifically for PET segmentation [22]. Furthermore, fully convolutional networks (FCNs) [23] have gained much attention for probability maps generation by extracting the high-level features of lesions and normal organs and producing coarse segmentation or bounding boxes to be used for tumor localization[24].

Figure 3 depicts a standard workflow for AI-based segmentation of PET and PET/CT images. These steps start with study design and data collection, and as we mentioned for supervised techniques, with manual delineations as the ground truth; PET/CT images need to be resampled to consistent sizes; considering a single- or bi-modal segmentation model, individual or multi-channel data should be prepared as input to the model. The evaluation on the test results to check the reliability the segmentation model should be optimally considered on the data from the independent centers that are captured under different conditions. The next step is he data and model sharing along with the list of limitations and negative results of the proposed model.

We can consider fully automated segmentation as a two-steps process that includes separate detection and segmentation modules [25, 26]. Lesion detection and segmentation can be performed simultaneously [26-28] or distinctly (back-to-back) as complementary tasks (using one model for detection [29] or two cascaded deep models for detection followed by segmentation [30]). Some existing segmentation techniques are designed based on the input from a detection step performed automatically [31] or manually [32] to localize the suspicious regions. In Table 1, a few studies that report detection performance, in addition to segmentation performance, are pointed out.

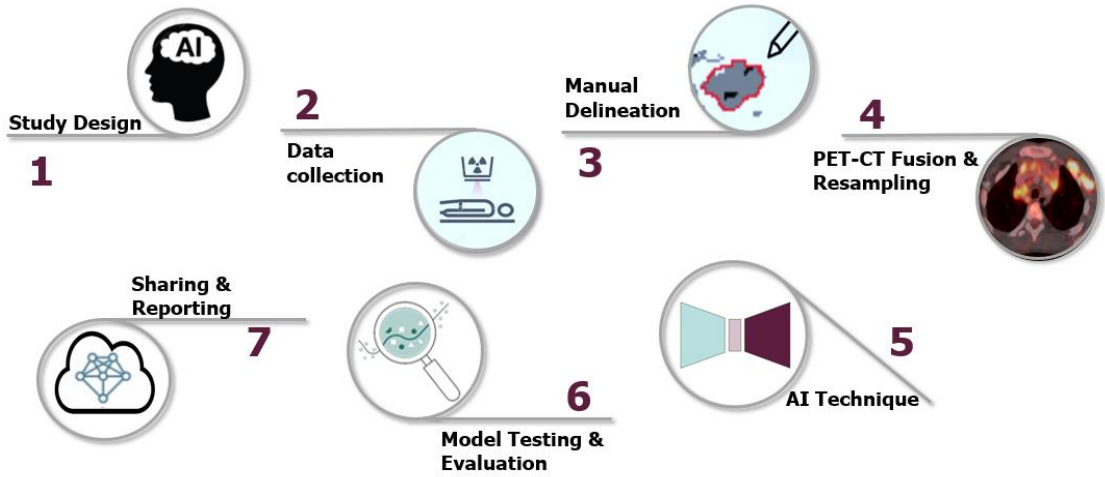

Figure 3: Standard workflow for AI-based segmentation in PET and PET/CT images, consisting of the following steps: (1) study design such as the need for automated segmentation for TMTV calculation or radiomics analysis, (2) PET or PET/CT data collection from relevant study cohort, (3) manual delineations provided by physicians or by using semi-automatic thresholding techniques to be used for training the supervised technique, (4) Data preparation including, cropping, resampling and data fusion considering the application, (5) Developing AI model (supervised/semi-supervised), (6) Model testing and evaluation the model to be applied on new data, (7) Sharing the model for transfer learning.

Table 1: Supervised PET/CT Segmentation Studies

| Authors | DP | Technique | Anatomic interest/dataset | Detection Performance | Segmentation Performance |
|---|---|---|---|---|---|
| Zhao et al. [33] | × | 3D FCN | Lung, 84 PET/CT | - | Dice=0.85 |
| Zhong et al. [24] | × | 3D U-Net + graph cut | Lung, 32 PET-CT | - | Dice (PET)=0.76<br>Dice (CT)=0.869 |
| Zhong et al. [34] | × | FCN | NSCLC, 60 PET/CT | - | Dice (CT)= 0.861 ± 0.037<br>Dice (PET) = 0.828 ± 0.087 |
| Li et al. [35] | × | FCN | NSCLC, 84 PET/CT | - | Dice=0.86±0.05<br>Sensitivity=0.86 ±0.07 |
| Perk et al. [36] | × | VGG19 | 14 NaF PET/CT | - | Accuracy=0.88<br>Sensitivity=0.9<br>Specificity=0.85 |
| Moe et al. [37] | × | U-Net | 197 H&N PET/CT | - | Dice (PET/CT)= 0.75 ± 0.12) |
| Zhao et al. [38] | × | FCN + auxiliary paths | 30 H&N PET-CT | - | Dice=0.8747 |
| Kumar et al. [26] | ✓ | CNN | 50 NSCLC PET-CT | Precision: 64.6 ± 29.61<br>Sensitivity: 80.0 ± 28.3<br>Specificity: 99.89 ± 0.13<br>Accuracy: 99.85 ± 0.14 | Dice= 0.6385 |
| Andrearczyk et al. [39] | × | 2D and 3D V-Net | 202 H&N PET/CT | - | 2D Dice (PET-CT)= 0.606<br>3D Dice= 0.597 |
| Iantsen et al. [40] | × | U-Net with Squeeze & Excitation Normalization | 254 H&N PET/CT | - | Dice=0.759<br>Precision=0.833<br>Recall=0.74 |
| Ma et al. [41] | × | CNN + hybrid active contours | 254 H&N PET/CT | - | Dice=0.752<br>Precision=0.838<br>Recall=0.717 |
| Yousefirizi et al. [42] | × | GAN+Mumford-Shah loss +ACM | 201 H&N PET/CT | - | Dice=0.82±0.06<br>Jaccard=0.81±0.07<br>HD=1.72±0.67 |
| Weisman et al. [30] | ✓ | DeepMedic | 90 lymphoma PET/CT | Sensitivity= 87% (3 fp) | Dice= 0.64 (interquartile range:0.43-0.76) |
| Li et al. [20] | × | DenseX-Net | 80 lymphoma PET/CT | - | Dice=0.728 |
| Jin et al. [43] | × | U-Net | 110 esophageal PET/CT | - | Dice=0.764 ± 0.134<br>HD=47 ± 56mm |
| DP: Detection Performance reported<br>NSCLC: Non-small cell lung cancer; H&N: Head and Neck | | | | | |

## 2.1. Supervised AI based attempts for PET-only Segmentation

Here we briefly describe some AI techniques used for PET segmentation. It is worth noting that performing accurate and reproducible tumor delineations on PET images is difficult due to partial-volume effects (PVEs), noise, motion artifacts, and varying shape, texture, and location of tumors [11, 13]. Variations in image properties due to varying PET/CT scanners in real clinical practice is also a challenge. Most conventional and AI-based techniques for segmentation are based on classification of each voxel in the PET image to tumor (or a specific normal organ) vs. background region, a task which is affected by these limitations. As an example, high repeatability of segmentation for smaller lesions in PET is hard to achieve since PVE affects the apparent tumor uptake [44].

Czakon et al. [45] applied different AI techniques for PET-only segmentation, namely 3D spatial distance weighted fuzzy c-means [46], dictionary based model [47] and CNN (e.g. 3D U-net architecture) for PET segmentation, and CNN showed better performance compared to other techniques. The superior performance of CNNs was also confirmed later by the first MICCAI challenge in 2018 on PET segmentation on a dataset composed of simulated, phantom and clinical scans [22]. CNNs have nowadays become very popular networks for PET segmentation; for instance Blanc-Durand et al. [27] and Huang et al. [48] also used 3D U-net for automated tumor segmentation in gliomas and head and neck cases respectively.

Most AI segmentation techniques in PET have been for FDG PET scans [27, 45] [48] [49], the other radiotracers have been rarely considered. As an example, Kostyszyn et al. [50] used a 3D U-net architecture to segment intra-prostatic tumors in PSMA PET scans. Zhao et al. [51] developed a 2.5D U-Net architecture for segmentation of prostate lesions, local and secondary prostate tumors in lymph nodes and bones. Iantsen et al. [25]applied an SE U-net for tumor detection and segmentation in PET images of cervical cancer cases. Their proposed technique is capable of differentiating the pathological and physiological uptake in bladder successfully. In any case, majority of works have involved both PET and CT for segmentation, which we describe next.

## 2.2. Supervised AI techniques for Tumor Co-Segmentation from PET/CT images

Personalized therapy decision can be guided by PET/CT since the corresponding voxels in PET and CT contain complementary but distinct information [34] [52]. Table 1 summarizes the main studies on co-segmentation of tumors in bi-modality PET and CT images. U-net [53], 3D U-net [54] and V-net [55], are widely used architectures for PET/CT segmentation. Recently, state-of-the-art frameworks such as skip connections [56], dense-net [57], recurrent residual convolutional neural network [58], GAN [59] along with integrating squeeze and excitation modules [60], Deepmedic [61] and attention mechanism [62] have gained much attention for PET/CT segmentation.

Multi-modality segmentation methods aim to utilize the functional information of PET images and anatomical localization of CT images simultaneously [26, 35, 43, 51] or separately [63, 64] [26]; thus PET/CT fusion is needed for PET/CT segmentation. For tumor segmentation, as lesions can spread throughout the body, spatially variant fusion techniques can improve segmentation performance [16]. PET/CT fusion for segmentation can be varied from using a multi-channel input (input level fusion) [36] to the layer level fusion the modality specific encoder branches [24, 34, 63, 65]. In the modality-specific framework, multi-channel input can consist of CT + PET, or CT + maximum intensity projection (MIP) PET images entered into the segmentation model.

CNN based PET/CT segmentations are mainly carried out based on image patches around the tumor without considering tumor occurrence in different parts of the images [24] [33] [35]. To cope with this limitation, a spatially varied fusion map proposed by Kumar et al. [26] measures the relative significance of PET and CT features in the different parts of the images. Their suggested co-learning scheme involves (i) a CNN that learns to extract the spatially varying fusion maps, and a (ii) fusion operation that prioritizes the features from each modality. Figure 4 shows the fusion technique proposed by Yuan et al. [66]. For DLBCL segmentation, they utilized two encoder branches for single-modality feature extraction and then used hybrid learning module in a supervised 3D CNN to creates a prediction map of DLBCL lesions.

On the other hand, some existing multi-modality PET-CT segmentation techniques are time-consuming or require pre-processing steps including clipping, standardization, and resampling (isotropic or anisotropic) for one or both modalities and post-processing steps [33]. Furthermore, the fact that these distinct modalities describe complementary but not identical characteristics of the same target is ignored in some studies [67-69].

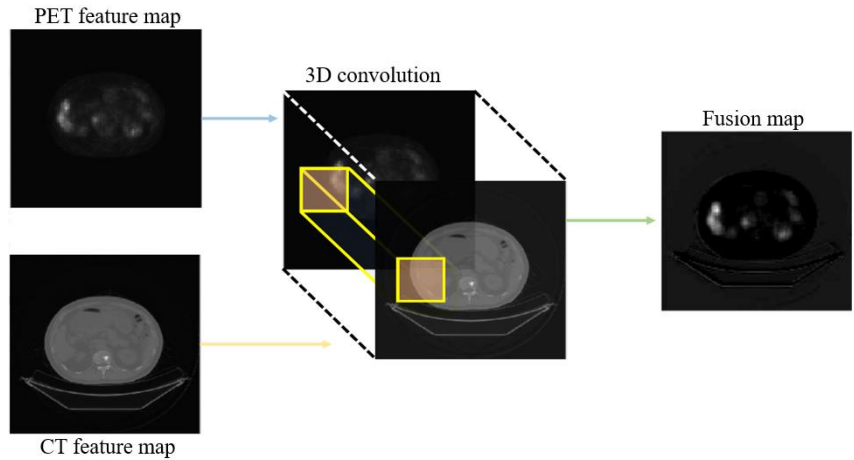

Figure 4: Conceptual description of hybrid learning generating fusion maps by 3D CNNs. The encoder branches extract features form PET and CT images. This process (an example shown above) is repeated for different feature maps via different layers. The spatial fusion maps are generated by hybrid learning that quantifies the contribution of the complementary information from PET and CT images. The learned feature maps are then concatenated (re-used with permission from Yuan et al. [66])

### 2.3. AI techniques for unlabeled data or data with scarce or weak annotations

The performance of AI-based techniques improves logarithmically with the size of training data [70] [71]. At the same time, consistency of the labels is of primary importance. As an example, Weisman et al. [29] showed that the detection performance of the Deepmedic [61] model will not improve after training with 40 or more patients. This can be explained as follows. Delineations by experts have in the past been mostly SUV-based, impacting reliability of supervised techniques. Limited number of annotations can be considered as the problem with "scarce annotations", i.e. labeled data is rare, while "weak annotations" occur when the existing labels are noisy or roughly drawn or inaccurate. Data scarcity emerges from the class imbalance of medical images and time-consuming task of manual delineations. Meanwhile limited consistency and reliability of annotations clearly affect AI task performance. These limitations motivates the use of advanced AI techniques that can be trained with limited supervision i.e. semi- supervised techniques [72, 73] [74, 75] and unsupervised [76] methods.

### 2.3.1. Unsupervised techniques

A fully unsupervised and reliable segmentation framework for PET/CT remains to be demonstrated. Here, we briefly consider some existing studies to this end. Unsupervised techniques based on clustering have shown acceptable performance for tumor segmentation in PET images considering heterogeneous uptake patterns and vague edges. Addressing the inherent imprecision of PET, Lian et al. [77] suggested using Dempster-Shafer theory, to model the uncertainty along with an evidential clustering algorithm considering the intensity of the voxels and textural features of a patch surrounding the voxel. The same group also proposed a belief function to model uncertain image information and an adaptive distance metric to consider the spatial information [69, 78]. Hu et al. [79] aggregated the voxels of 3D PET scans to supervoxels ( a cluster of voxels) and subsequently used density-based spatial clustering with noise (DBSCAN) for segmentation.

Recently unsupervised AI techniques have been applied for anomaly detection in medical images based on normal images. The idea of employing normal images (without anomalies) to train unsupervised anomaly detection and segmentation models has gained much attention mostly using encoder-decoder or CNNs [80]. For example, training a convolutional adversarial auto-encoder on normal images can be used to learn a latent space that models the variant normal PET images. The residual map is then calculated to identify the PET images that are different from this manifold. Wu et al. [81] applied this technique on lung cancer images and their method outperformed U-net. Klyuzhin et al. [82] used this idea for background removal to predict the physiological PSMA-PET (18F-DCFPyL) uptake patterns from a pair of CT and low-resolution PET images.

In a nutshell, as unsupervised learning techniques are less affected by the quality of labeled data which may be inadequate and biased with limited diversity. These approaches have the potential for flexible application to heterogeneous patient data, even for rare diseases, and can be combined with expert interpretations.

### 2.3.2. Weakly-supervised and semi-supervised techniques

Weakly supervised learning is a general term for training schemes under incomplete, inexact and inaccurate supervision [14] as defined in section 1. Afshari et al. [72] proposed a weakly supervised technique using FCN with modified Dice and Mumford-Shah loss functional for tumor segmentation in PET images of head and neck, while only the bounding boxes around the tumors (weak annotations) were provided.

Semi-supervised learning as a sub-category of weakly supervised learning techniques use unlabeled data along with a limited number of labeled data for training. For instance, in the joint training strategy proposed by Li et al. [20], the network parameters and the labels for unlabeled data were iteratively updated. During training, the optimal convolutional kernel is determined that improves the accuracy of the segmentation. The authors used this parallel segmentation and reconstruction flows for lesion segmentation in DLBCL PET/CT images.

### 2.4. Estimation-based approaches

PVEs in PET arise mainly due to two reasons: limited system resolution and finite voxel size [83] [84]. The latter results in tissue-fraction effects (TFE), i.e. a voxel containing more than a single tissue. Conventional segmentation methods[1], including deep-learning-based approaches, classify each voxel as belonging to only one tissue type, and thus have limited efficacy in addressing the TFE. To address this inherent limitation, recently, techniques have been proposed that, estimate the volume that a tissue occupies within a voxel using an encoder-decoder network, which can then be used to define a segmentation [67, 85]. The methods have shown improved accuracy compared to conventional methods, including U-net-based methods, on the task of segmenting tumors in patients with non-small cell lung cancer, as demonstrated in a study with the ACRIN 6668/RTOG 0235 multi-center clinical trial data [67]. Further, these approaches can also use training data derived from other modalities where the resolution may be higher [85] (Figure 5).

### 2.5. Segmentation of normal organs

The occurrence of abnormalities (tumors and metastasis) can be very unpredictable and heterogeneous, while the spatial information of normal organs with physiological uptakes are relatively stable, consequently segmentation of normal organs from PET/CT can be a preliminary step for automatic tumor segmentation. Furthermore, the diverse size, variant shape and unpredictable location of metastasis occurrences, impose the need for diverse delineated images to achieve the generalizability and good performance of the segmentation model [82].

---

[1] The term "conventional segmentation techniques" in this paper refers to thresholding methods, region-growing methods and statistical methods (i.e. as opposed to AI techniques ). By contrast, "conventional segmentation techniques" in the literature sometimes refers to SUV-based thresholding techniques that is only a subset of what we denote by this term.

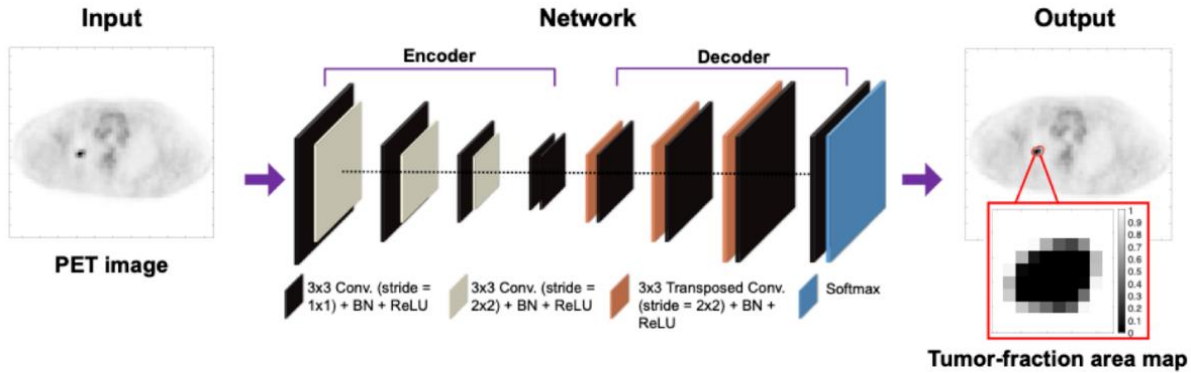

Figure 5: The Bayesian approach proposed by Liu et al. [67] to tissue-fraction estimation for oncological PET segmentation. Adapted from [67].

Normal organ detection can be applied to determine reference level uptakes to help define the Deauville 5-point scale in lymphoma cases; i.e. in reference to mediastinal blood pool and liver. Sadik et al. [86] trained an FCN to extract the liver and the mediastinal blood pool using CT images. Many normal organ segmentation approaches rely solely on CT e.g. using DL models [87] [88] specifically there are publicly share annotated CT images [89]. Yu et al. [90] segmented the normal organs based on CT images by applying a multi-atlas method and removed them to obtain lymphoma lesions [79, 90]. However, removing the organs that are considered as "normal" on CT images, may not take into account the possible abnormal uptakes that are observed in these organs. On the other hand, PET-based segmentation of normal organs can be challenging depending on the radiopharmaceutical with possible very low normal organ uptake. The existing studies are mostly based on PET/CT and the corresponding CT images provide the anatomical reference [82].

Seifert et al. [19] described a semi-automatic approach to distinguish normal organ uptake and tumor uptake in PSMA PET/CT imaging by applying a GAN following SUV thresholding to segment a range of normal organs on patient CT images and excluding the regions with physiological PSMA uptake [19, 82]. Recently, Klyuzhin et al. [91] applied a set of U-net models to perform segmentation of each normal organ.

## 3. Solutions to Tackle Limitations in Annotations

As we previously mentioned, the need for relatively large and consistent labeled data [92], in spite of the scarcity of labeled data in the field of medical imaging, should be addressed in order to develop reliable and generalizable AI techniques for segmentation. Based on the recommendations by the task group 211 of the AAPM (American Association of Physicists in Medicine), thorough, consistent and sufficient evaluation of developed PET automatic segmentation method should be applied on i) phantom images, ii) a combination of physical and numerically simulated phantom images, and iii) clinical images [13]. These recommendations arise from the fact that the volume of clinical images available for training and evaluation of segmentation models is often limited [13].

### 3.1. Data augmentation

Data augmentation (i.e. flipping, shifting, rotating and random cropping) is a preliminary solution to tackle the lack of labeled data. However, these augmentation techniques sometimes produce meaningless medical images; consequently, realistically simulated or synthesized images can be used in this regard. GANs, for instance, can leverage data augmentation by synthesizing realistic-looking PET images based on existing PET or CT images. Studies on synthesizing PET images based on CT data have mostly employed conditional GANs (cGANs)[93] and multi-channel GANs [94, 95]. Ben-Cohen et al. [96] used a cGAN to synthesize liver PET images based on CT images. cGANs can produce very realistic PET images but tumor regions are not very well reproduced [97]. On the other hand, FCNs have shown promising results for tumor synthesis but the synthesized images are blurry. Consequently, using a combination of cGANs and FCNs showed improved performance [96]. Multi-channel GANs (mGANs) was also applied to generate PET images based on CT and label images [94].

### 3.2. Using simulation studies for training

Using simulations to generate PET images with known tumor boundaries provides a way to address the challenge of limited accessibility of annotation. Briefly, an anthropomorphic digital phantom population, consisting of the tumors with known ground-truth boundaries, can be used to generate PET images using tools that model the PET imaging physics. There exist multiple mechanisms to generate such images. One approach is to use anthropomorphic phantoms, such as those based on XCAT [98, 99] as inputs to software that model the PET physics. A second approach is synthetically generating tumors and inserting into clinical images. For example, Leung et al [100] used a stochastic approach to simulate tumors with parameters such as shape/size/uptake similar to those in clinical images, and used a projection-domain approach to insert these tumors into clinical images. Figure 6 depicts their proposed method. They demonstrated that pre-training a network using this data led to improved segmentation accuracy and lower requirements of training-data compared to a network that only used clinical data for training. Further, the method was relatively insensitive to PVEs and generalized across scanners. In using simulation-based strategies, it is important that the simulations be realistic, especially in terms of modeling the clinical characteristics. To evaluate this realism, observer-based strategies can be used [101].

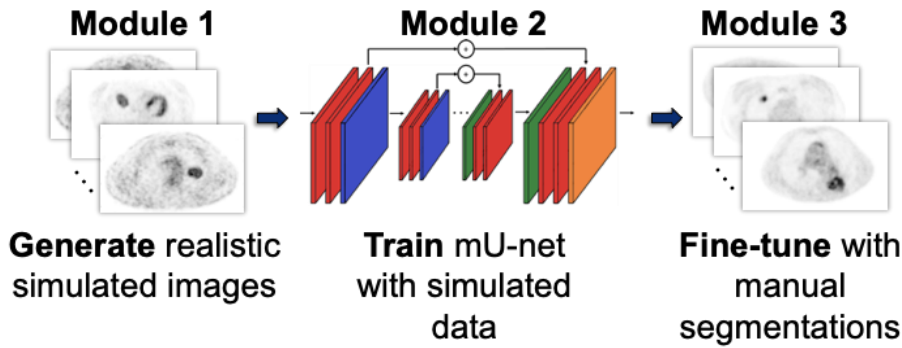

Figure 6: Simulation-based segmentation framework proposed by Leung et al. [100] (re-used with permission).

### 3.2.1. The consistency of ground truth

Annotations for standard training data have generally high cost and are laborious and time consuming for clinically experienced professionals. The delineations are made for different goals: for instance, masks generated for radiotherapy planning are generally larger than the original size of the tumors [102, 103]. Depending on the interest in shell, surrounding tissue and tumor region, delineations can vary. For instance, there has been increasing interest in peri-tumoral radiomics analyses, moving beyond typical delineated boundaries of tumors [104, 105]. Besides, there are no actual hard edges between tumor and surrounding tissues even at the microscopic level. Furthermore, delineations based on fixed thresholding may be especially prone to error due to PVEs. Semi-automated algorithms can help obtain more consistent and accurate ground truths such Fuzzy locally Adaptive Bayesian (FLAB) [106].

Majority voting between manual segmentation by different experts has been used to enhance consistency. Web-based or freely available tools for crowdsourcing can also be used based on labels made by untrained or non-expert trainees as a cost-effective alternative especially for organ segmentations [71, 107]. Mehta et al. [107] showed that the performance of their 3D U-net model for kidney segmentation trained on crowdsourcing labels was not significantly different compared to expert-labeled. Heim et al. [108] used majority voting on labels for liver segmentation; they showed that crowd segmentation matched expert segmentation. There are strong observations that a large collection of naïve but independent analyses can outperform individual performance even by experts [109]. Furthermore, techniques have been developed to underemphasize the delineations made by lower-expertise [72].

## 4. Evaluation of AI techniques

Conventionally, segmentation methods are evaluated by comparing the estimated segmentation with the gold standard (ground truth) segmented masks, and quantifying performance by some measure of distance. For evaluating segmentation techniques, the use of at least three metrics is recommended since some of them are correlated (i.e. Dice score (DSC), Jaccard score (JSC), Hausdorff distance (HD), average distance (AVD) and the Mahalanobis distance (MHD) are highly correlated) [110]; for instance Dice, true positive rate (TPR) and false negative rate (FNR) can be used [111]. Based on the recommendations by AAPM TG 211, positive predictive value (PPV) and sensitivity should also be considered. For radiotherapy planning applications, sensitivity evaluation should be preferred, whereas for radiomics/quantification purposes, PPV is more informative. (A number of frequently invoked metrics are listed in Table 2).

Table 2: Evaluation criteria for segmentation

| Evaluation measure | Definition |
|---|---|
| Dice score (DSC) | $\frac{2TP}{2TP + FP + FN}$ |
| Jaccard (JSC) | $\frac{TP}{TP + FP + FN}$ |
| Hausdorff distance (HD) | $max\{sup_{x \in X} d(x, Y), sup_{y \in Y} d(X, y)\}$ |
| Sensitivity | $\frac{TP}{TP + FN}$ |
| Specificity | $\frac{TN}{TN + FP}$ |
| Positive predictive value (PPV) or precision | $\frac{TP}{TP + FP}$ |

TP=True Positive, TN= True Negative, FP= False Positive, FN= False Negative.

Clearly, this evaluation metrics can be affected by the ground truth accuracy are not defined based on the ultimate task that segmentation is performed for; in this section we consider the no gold standard evaluation [112] that has been proposed to tackle this problem.

State-of-the-art DL segmentation models remain to be broadly translated to routine clinical workflow [113]. Such a high bar requires tacking of significant challenges with generalizability, repeatability, reproducibility and trustworthiness of AI-based segmentation techniques, which we describe next.

### 4.1. No-gold-standard evaluation

Conventionally, segmentation methods are evaluated by comparing the estimated segmentation with the ground-truth segmentation and quantifying performance by some measure of distance, such as dice scores and Jaccard scores that quantify spatial overlap, or Hausdroff distance that quantifies shape similarity. However, this strategy suffers from two issues. The first issue is the lack of such ground-truth segmentations. While manually defined segmentations serve as surrogate ground truth, they can be erroneous, suffer from inter and intra-reader variability and be difficult to obtain (time, expense). A second issue is that medical images, including PET images, are segmented for tasks such as quantification. Evaluation methodologies that quantify the distance between the measured and surrogate ground-truth segmentations may not correlate with that task [114].

A segmentation method developed to measure a certain quantitative feature from an image so it should be evaluated based on how well the method performs on the task of reliably measuring that feature. However, this evaluation requires knowledge of the true quantitative parameter, or some gold-standard measurement. However, that is often unavailable or difficult to obtain. To address this issue, no-gold-standard evaluation (NGSE) techniques have been proposed [112, 115-118], including in the context of evaluating PET segmentation methods on the task of estimating MTV [119] [120] and most recently, on evaluating PET partial volume compensation methods on the task of measuring activity uptake [121].

Given measurements from multiple quantitative imaging methods, the NGSE techniques assume a linear relationship between the true quantitative values and the quantitative values obtained with each imaging method. This relationship is parameterized by a slope, bias, and a Gaussian distributed noise term described by a standard deviation. As we would expect, estimating these terms can yield a measure of how reliably the quantitative values are estimated (see Fig. 1 in [119]). Next, assuming that the true values are sampled from a certain parametric distribution, the NGSE technique derives a statistical model of the measurements obtained with the different imaging methods. The technique then estimates the parameters of the linear relationship that maximize the probability of occurrence of these measurements. The ratio of the noise standard deviation and slope terms (NSR) for each method are then used to rank the methods based on how precisely they measure the true quantitative value. As has been shown in multiple studies [112, 115-118], this technique is able to accurately rank different quantitative imaging methods based on how precisely these methods measure the true value. For example, in Figure 7, we show the performance of the NGSE technique in ranking three different quantitative SPECT methods. We observe that the NGSE technique, even in the absence of ground truth, yielded the same ranking as when the ground truth was known.

This NGSE technique promises to address a major barrier with clinical evaluation of segmentation methods for PET, but several challenges need to be addressed. As an example, existing NGSE techniques assume that the noise between the different methods is correlated. However, Liu et al recently proposed a strategy to model correlated noise [120]. Another challenge is that NGSE techniques may require large amounts of patient images (N > 200). A Bayesian approach to reduce the number of patient studies needed by the NGSE technique has demonstrated promise to address this issue [122]. Overall, these ongoing studies provide promise that NGSE techniques are well poised to provide a mechanism for clinical evaluation of PET segmentation methods on quantitative tasks.

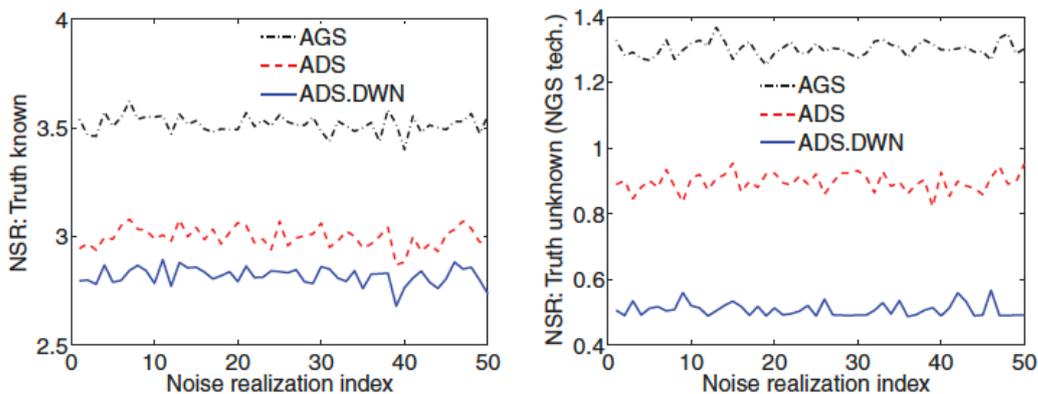

Figure 7: Results from a study showing the performance of the NGSE technique. In this study, three reconstruction methods for quantitative SPECT (AGS, ADS and ADS.DWN) were evaluated on the task of measuring regional uptake

values. The figure on the left shows the rankings of the methods on the task of precisely measuring the true values, as quantified by the noise-to-slope ratio (NSR), that were obtained when the true uptake values were known. The figure on the right shows the results using the NGSE technique that did not use the knowledge of the true uptake values. The experiment was repeated for 50 noise realizations. We observe that even in the absence of the ground truth, the NGSE technique yielded the same ranking as the rankings obtained when the ground truth values were known (Figure adapted from [112]).

## 4.2. Generalizability

Most AI techniques for medical imaging suffer from i) the bias in the training population, ii) data leakage (the test data are not actually "un-seen" in these studies) or iii) overfitting that results in the lack of generalizability of the AI models [123]. AI techniques should be evaluated on data from different scanners, centers and patient populations with different clinical characteristics and demographics [100, 124]. The cross-center generalizability of the model can be considered in two ways: 1) Training the segmentation model on the data from one center (scanner) and testing on data from another center. 2) Training the model on data from one center (scanner) and testing on data from other centers (scanners) [100]. The leave-one-center-out cross-validation can also help the generalizability [39, 42], in which, one center can be taken as the test set and the remaining centers as the training set. AI techniques for automatic segmentation have been mainly tested on limited dataset (in terms of the number of data and heterogeneity. Consequently semi- automatic thresholding methods remain as the main segmentation method in practice [44].

## 4.3. Repeatability and Reproducibility

Repeatability measures the amount of variable results in repeated evaluations with same data, scanner type, reconstruction algorithm, as well as image noise levels. On the other hand, reproducibility considers the variability of the results when one or more of the above-mentioned conditions are variable. There are very limited studies that consider the repeatability of AI segmentation techniques; while many more studies have reported the reproducibility of them. A repeatable segmentation technique produces comparable results on test-retest images (PET or PET/CT) of the same patient with similar physiological conditions [125], Pfaehler et al. [44] considered the repeatability of two segmentation techniques, i.e. U-net framework and textural feature + random forest classifier. They used a fully independent test–retest dataset of 10 PET/CT NSCLC recorded on two consecutive days. They concluded that AI-based segmentation approaches have shown better repeatability compared to conventional segmentation methods [126].

Reproducibility analysis can be performed by intra-class correlation (ICC) that helps to compare intra- and inter-individual variabilities. For repeatability analysis, to assess within-subject variability under identical conditions [127], percent test-retest differences in PET imaging are often quantified. For visual assessment of the mean versus differences of test-retest observations, Bland-Altman plots are also very helpful [128]. The reader is especially referred to review article by Lodge [129] to understand the links between different metrics to quantify repeatability.

Furthermore, we note that there are a variety of evaluation metrics for segmentation, and multiple metrics may need to be utilized for more thorough assessment. For instance, a metric (e.g. Dice, Jaccard) can consider the intersection of the predicted mask and ground truth, while it is not able to consider the edge details of the predicted mask achieved by other metrics (e.g. Hausdorff distance) [110].

In manual or semi-automatic delineations, inter-observer variability for delineations refers to the different segmentations obtained by different physicians while intra-observer variability is for segmentations made by a physician at different instances. For automated AI methods, e.g. DL methods, variabilities for segmentation originate from i) the inherent variability of dataset, ii) random initialization of network parameters, ii) the stochastic optimization process, iv) variable selection of hyper-parameters, and v) variability of the infrastructure. These variabilities challenge high reproducibility [111]. To tackle this, a number of recommendation have been made. As shown in Figure 8, this includes providing a) adequate descriptions for the DL frameworks as well as b) analysis of variability due to different factors, arriving at c) overall analysis for the sources of variability, towards efficient evaluation of segmentation results [111]. There are also a number of CHECKLISTS to this aim (CLAIM, Checklist for Artificial Intelligence in Medical imaging [130]).

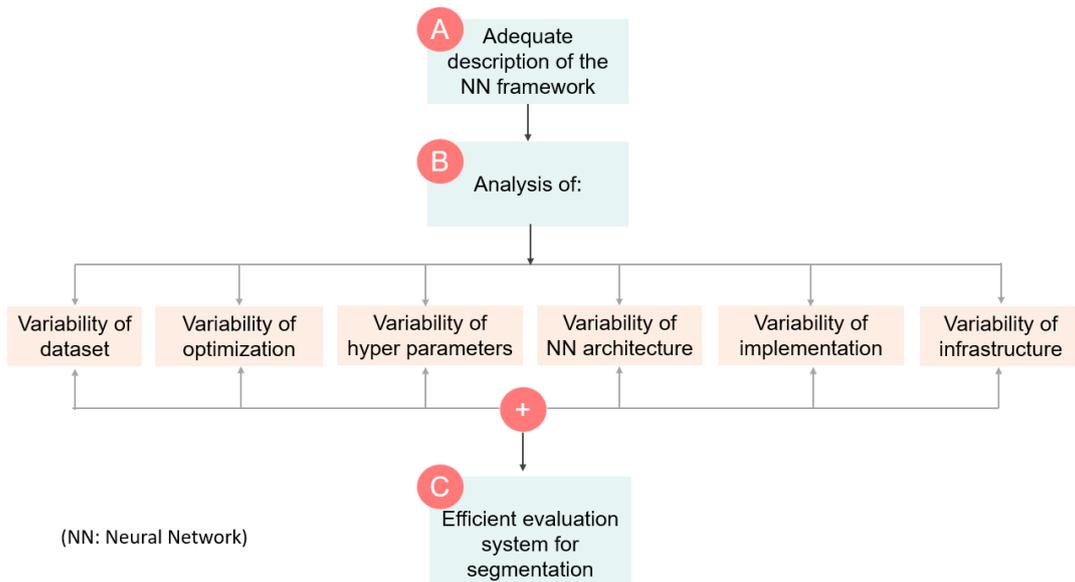

Figure 8: Three main recommendations (A, B and C) to address the potential issues with reproducibility of DL frameworks for medical image segmentation (re-used with permission [111]).

### 4.4. Trustworthiness

AI techniques should be designed to allow physicians make better decisions, fostering their autonomy and minimizing automation bias ('human agency') and preserve the security and privacy of patients. Transparency is a crucial component of trustworthiness for an AI technique, which is challenged by "blackbox" models.

The performance of AI-techniques and the demand to use them in clinics can be improved if AI systems are able to represent uncertainties in given tasks. Two main categories of uncertainty are (i) epistemic uncertainty that refers to the uncertainty in the model, and (ii) aleatoric uncertainty that addresses the noise or randomness [131] and the spatial transformation of the input images [132]. These uncertainties can be considered by AI techniques such as Baysian dilated CNN to predict the segmentation and generate corresponding spatial uncertainty map. Deep ensembles was also suggested to estimate the uncertainty as the variance of predictions by multiple models [133].

The segmented regions with high uncertainty can be referred to radiologists this helps to improve the quality of decision systems based on AI techniques. Consequently, radiologists can evaluate this uncertainty as a "human-in-the-loop" setting to improve the segmentation performance. This setting also help to suppress implausible segmentations that are impossible to be produced by a radiologist [113, 134].

The performance monitoring of AI techniques when applied on real data in clinics should also be iteratively evaluated for continual improvement. It is also recommended that safety and robustness of AI techniques should be evaluated based on the updated methodology [135].

## 5. Further Advancements

Scarce annotations, as previously mentioned, result in supervised segmentation models with limited generalizability. To tackle these, a number of newer methods, beyond above-mentioned efforts, are also being developed and explored, which we briefly discuss next.

## 5.1. Federated learning (FL)

AI techniques for segmentation have faced challenges with generalizability due to limited volumes of data, having heterogeneity and variability in size, shape, texture, and site of the pathologies [111]. FL enables training a centralized AI model across institutions instead of transferring the de-identified data from different centers to the centralized data storage. Parallel and sequential approaches have been proposed. In parallel training, the data is divided and different models are trained on each part, and the weights are transferred to the central model [136, 137]. In sequential approach, the model is trained on the data from each institution and cyclic weight transfer is applied [138].

FL faces a number of limitations and challenges. Data from different institutions with different infrastructure, imaging workflows and different standards for delineations are heterogeneous with limited scalability. There may also exist ambiguity in pre-processing steps that are done differently in each institution. Finally, although only weights are supposed to be shared in FL, leakage of sensitive information is still possible which needs to be tackled [71, 139].

## 5.2. Self-training techniques

As we discussed earlier, semi-supervised learning techniques utilize the information of unlabeled data to train the model when limited and/or weakly annotated data are available [140]. Self-training is a semi-supervised technique increasingly explored to estimate labels for unlabeled data during training. The predicted "pseudo-labels" are weighted based on their confidence and then concatenated with labeled data to re-train the network by augmented training data. Self-training techniques remain to be thoroughly explored towards PET segmentation. As a limitation, incorrect early predictions can be propagated back to the network during the training process [141], though this issue can be suppressed by techniques such as curriculum learning [142]. Curriculum learning that uses regression networks to predict the target region can overcome this limitation by enforcing the predictions of the unlabeled images to be close to the inferred label (in terms of target size or global label distributions). The regression step can regularize the segmentation model and reduce the errors of predicted pseudo labels for the unlabeled data.

## 5.3. Neuro-Symbolic AI models for PET segmentation

Clinicians refer to a set of conditions, different imaging modalities, patient treatment and surgery history, and biological and physiological conditions to evaluate a suspicious lesion based on their experiences [143]. By incorporating data, images as well as "rules" into the AI models for detection and segmentation, more accurate and reliable AI based tasks are expected. As such, use of neuro-symbolic neural networks and explainable AI techniques [144] may add significant value. Overall, the future of AI may lie in the bringing together of two historically distinct and divergent paradigms in AI (symbolic vs. connectionist). Recently, use of hypertexts [145] and interactive reporting [71] have been considered to extract rules, towards generation of more generalizable models from limited data.

## 6. Conclusion

Segmentation is a vital task for MTV calculation, radiotherapy planning and radiomics analysis. Although AI techniques have significant potential for automated segmentation of oncology PET and PET/CT images, major challenges remain in terms of lack of consensus for manual delineations, and inter-/intra-observer variabilities, to build consistent training sets for supervised techniques. It has indeed been shown that for supervised AI techniques, consistency in the training data is of higher importance compared to access to large amounts of data [30]. Meanwhile, to tackle issues with limited (annotated) data, a number of approaches such as semi supervised, self-training framework, federated learning and neuro-symbolic AI are being actively explored. For an AI-based segmentation technique to translate to routine clinical workflow, significant efforts are needed for improved generalizability and trustworthiness. Overall, AI-based segmentation methods applied to oncological PET images hold significantly promise and potential to help enable personalization of therapy for cancer patients.

**Acknowledgements**

This project was in part supported by the Natural Sciences and Engineering Research Council of Canada (NSERC) Discovery Grant RGPIN-2019-06467, and the Canadian Institutes of Health Research (CIHR) Project Grant PJT-173231. We would also like to acknowledge Dr. Ghassan Hamarneh and Kumar Abhishek from Simon Fraser University for very valuable discussions.

REFERENCES

1. Langlotz, C.P., B. Allen, B.J. Erickson, et al., *A roadmap for foundational research on artificial intelligence in medical imaging: from the 2018 NIH/RSNA/ACR/The Academy Workshop.* Radiology, 2019. **291**(3): p. 781-791.
2. Litjens, G., T. Kooi, B.E. Bejnordi, et al., *A survey on deep learning in medical image analysis.* Medical image analysis, 2017. **42**: p. 60-88.
3. Vial, A., D. Stirling, M. Field, et al., *The role of deep learning and radiomic feature extraction in cancer-specific predictive modelling: a review.* Transl Cancer Res, 2018. **7**(3): p. 803-816.
4. Zwanenburg, A., M. Vallières, M.A. Abdalah, et al., *The image biomarker standardization initiative: standardized quantitative radiomics for high-throughput image-based phenotyping.* Radiology, 2020. **295**(2): p. 328-338.
5. Starmans, M.P., S.R. van der Voort, J.M.C. Tovar, et al., *Radiomics: data mining using quantitative medical image features*, in *Handbook of Medical Image Computing and Computer Assisted Intervention*. 2020, Elsevier. p. 429-456.
6. Klyuzhin, I.S., M. Gonzalez, E. Shahinfard, et al., *Exploring the use of shape and texture descriptors of positron emission tomography tracer distribution in imaging studies of neurodegenerative disease.* Journal of Cerebral Blood Flow & Metabolism, 2016. **36**(6): p. 1122-1134.
7. van Velden, F.H., G.M. Kramer, V. Frings, et al., *Repeatability of radiomic features in non-small-cell lung cancer [18 F] FDG-PET/CT studies: impact of reconstruction and delineation.* Molecular imaging and biology, 2016. **18**(5): p. 788-795.
8. Guezennec, C., D. Bourhis, F. Orlhac, et al., *Inter-observer and segmentation method variability of textural analysis in pre-therapeutic FDG PET/CT in head and neck cancer.* PloS one, 2019. **14**(3): p. e0214299.
9. Yang, F., G. Simpson, L. Young, et al., *Impact of contouring variability on oncological PET radiomics features in the lung.* Scientific reports, 2020. **10**(1): p. 1-10.
10. Caldwell, C.B., K. Mah, Y.C. Ung, et al., *Observer variation in contouring gross tumor volume in patients with poorly defined non-small-cell lung tumors on CT: the impact of 18FDG-hybrid PET fusion.* International Journal of Radiation Oncology* Biology* Physics, 2001. **51**(4): p. 923-931.
11. Foster, B., U. Bagci, A. Mansoor, et al., *A review on segmentation of positron emission tomography images.* Computers in biology and medicine, 2014. **50**: p. 76-96.
12. Hansen, S., S. Kuttner, M. Kampffmeyer, et al., *Unsupervised supervoxel-based lung tumor segmentation across patient scans in hybrid PET/MRI.* Expert Systems with Applications, 2021. **167**: p. 114244.
13. Hatt, M., J.A. Lee, C.R. Schmidtlein, et al., *Classification and evaluation strategies of auto-segmentation approaches for PET: Report of AAPM task group No. 211.* Medical physics, 2017. **44**(6): p. e1-e42.
14. Zhou, Z.-H., *A brief introduction to weakly supervised learning.* National science review, 2018. **5**(1): p. 44-53.
15. Cheung, H. and D. Rubin, *Challenges and opportunities for artificial intelligence in oncological imaging.* Clinical Radiology, 2021.
16. Hatt, M., C. Cheze-le Rest, A. Van Baardwijk, et al., *Impact of tumor size and tracer uptake heterogeneity in 18F-FDG PET and CT non–small cell lung cancer tumor delineation.* Journal of Nuclear Medicine, 2011. **52**(11): p. 1690-1697.

17. Im, H.-J., T. Bradshaw, M. Solaiyappan, et al., *Current Methods to Define Metabolic Tumor Volume in Positron Emission Tomography: Which One is Better?* Nuclear medicine and molecular imaging, 2018. **52**(1): p. 5-15.
18. Taghanaki, S.A., K. Abhishek, J.P. Cohen, et al., *Deep semantic segmentation of natural and medical images: a review.* Artificial Intelligence Review, 2021. **54**(1): p. 137-178.
19. Seifert, R., M. Weber, E. Kocakavuk, et al., *Artificial Intelligence and Machine Learning in Nuclear Medicine: Future Perspectives.* Seminars in Nuclear Medicine, 2021. **51**(2): p. 170-177.
20. Li, H., H. Jiang, S. Li, et al., *DenseX-net: an end-to-end model for lymphoma segmentation in whole-body PET/CT images.* IEEE Access, 2019. **8**: p. 8004-8018.
21. Zhou, T., S. Ruan, and S. Canu, *A review: Deep learning for medical image segmentation using multi-modality fusion.* Array, 2019. **3**: p. 100004.
22. Hatt, M., B. Laurent, A. Ouahabi, et al., *The first MICCAI challenge on PET tumor segmentation.* Medical image analysis, 2018. **44**: p. 177-195.
23. Long, J., E. Shelhamer, and T. Darrell. *Fully convolutional networks for semantic segmentation.* in *Proceedings of the IEEE conference on computer vision and pattern recognition.* 2015. Boston, MA, USA.
24. Zhong, Z., Y. Kim, L. Zhou, et al. *3D fully convolutional networks for co-segmentation of tumors on PET-CT images.* in *2018 IEEE 15th International Symposium on Biomedical Imaging (ISBI 2018).* 2018. Washington, DC, USA: IEEE.
25. Iantsen, A., M. Ferreira, F. Lucia, et al., *Convolutional neural networks for PET functional volume fully automatic segmentation: development and validation in a multi-center setting.* European Journal of Nuclear Medicine and Molecular Imaging, 2021: p. 1-13.
26. Kumar, A., M. Fulham, D. Feng, et al., *Co-learning feature fusion maps from PET-CT images of lung cancer.* IEEE Transactions on Medical Imaging, 2019. **39**(1): p. 204-217.
27. Blanc-Durand, P., A. Van Der Gucht, N. Schaefer, et al., *Automatic lesion detection and segmentation of 18F-FET PET in gliomas: a full 3D U-Net convolutional neural network study.* PLoS One, 2018. **13**(4): p. e0195798.
28. Zhu, Z., D. Jin, K. Yan, et al. *Lymph Node Gross Tumor Volume Detection and Segmentation via Distance-Based Gating Using 3D CT/PET Imaging in Radiotherapy.* in *International Conference on Medical Image Computing and Computer-Assisted Intervention.* 2020. Lima, Peru: Springer.
29. Weisman, A.J., M.W. Kieler, S.B. Perlman, et al., *Convolutional Neural Networks for Automated PET/CT Detection of Diseased Lymph Node Burden in Patients with Lymphoma.* Radiology: Artificial Intelligence, 2020. **2**(5): p. e200016.
30. Weisman, A., M. Kieler, S. Perlman, et al. *Automated Quantification of Lymphoma On FDG PET/CT Images Using Cascaded Convolutional Neural Networks.* in *MEDICAL PHYSICS.* 2019. WILEY 111 RIVER ST, HOBOKEN 07030-5774, NJ USA.
31. Andrearczyk, V., V. Oreiller, and A. Depeursinge, *Oropharynx detection in PET-CT for tumor segmentation*, in *Irish Machine Vision and Image Processing*. 2020: Sligo, Ireland. p. Pp. 109-112.2020.Sligo, Ireland.
32. Weisman, A.J., M.W. Kieler, S. Perlman, et al., *Comparison of 11 automated PET segmentation methods in lymphoma.* Physics in Medicine & Biology, 2020. **65**(23): p. 235019.
33. Zhao, X., L. Li, W. Lu, et al., *Tumor co-segmentation in PET/CT using multi-modality fully convolutional neural network.* Physics in Medicine & Biology, 2018. **64**(1): p. 015011.
34. Zhong, Z., Y. Kim, K. Plichta, et al., *Simultaneous cosegmentation of tumors in PET-CT images using deep fully convolutional networks.* Med Phys, 2019. **46**(2): p. 619-633.
35. Li, L., X. Zhao, W. Lu, et al., *Deep learning for variational multimodality tumor segmentation in PET/CT.* Neurocomputing, 2019.
36. Bradshaw, T., T. Perk, S. Chen, et al., *Deep learning for classification of benign and malignant bone lesions in [F-18] NaF PET/CT images.* Journal of Nuclear Medicine, 2018. **59**(supplement 1): p. 327-327.
37. Moe, Y.M., A.R. Groendahl, O. Tomic, et al., *Deep learning-based auto-delineation of gross tumour volumes and involved nodes in PET/CT images of head and neck cancer patients.* European Journal of Nuclear Medicine and Molecular Imaging, 2021: p. 1-11.
38. Zhao, L., Z. Lu, J. Jiang, et al., *Automatic nasopharyngeal carcinoma segmentation using fully convolutional networks with auxiliary paths on dual-modality PET-CT images.* Journal of digital imaging, 2019. **32**(3): p. 462-470.

39. Andrearczyk, V., V. Oreiller, M. Vallieres, et al., *Automatic Segmentation of Head and Neck Tumors and Nodal Metastases in PET-CT scans*, in *Medical Imaging with Deep Learning MIDL*. 2020: Montreal.2020.Montreal.
40. Iantsen, A., D. Visvikis, and M. Hatt. *Squeeze-and-excitation normalization for automated delineation of head and neck primary tumors in combined PET and CT images*. in *3D Head and Neck Tumor Segmentation in PET/CT Challenge*. 2020. Lima, Peru: Springer.
41. Ma, J. and X. Yang. *Combining CNN and hybrid active contours for head and neck tumor segmentation in CT and PET Images*. in *3D Head and Neck Tumor Segmentation in PET/CT Challenge*. 2020. Lima, Peru: Springer.
42. Yousefirizi, F. and A. Rahmim. *GAN-based bi-modal segmentation using mumford-shah loss: Application to head and neck tumors in PET-CT images*. in *First Challenge, HECKTOR 2020, Held in Conjunction with MICCAI 2020*. 2020. Lima, Peru: Springer.
43. Jin, D., D. Guo, T.-Y. Ho, et al. *Accurate esophageal gross tumor volume segmentation in pet/ct using two-stream chained 3d deep network fusion*. in *International Conference on Medical Image Computing and Computer-Assisted Intervention*. 2019. Springer.
44. Pfaehler, E., L. Mesotten, G. Kramer, et al., *Repeatability of two semi-automatic artificial intelligence approaches for tumor segmentation in PET.* EJNMMI research, 2021. **11**(1): p. 1-11.
45. Czakon, J., F. Drapejkowski, G. Zurek, et al., *Machine learning methods for accurate delineation of tumors in PET images.* arXiv preprint arXiv:1610.09493, 2016.
46. Guo, Y., K. Liu, Q. Wu, et al., *A new spatial fuzzy c-means for spatial clustering.* Wseas Transactions on Computer, 2015. **14**: p. 369-381.
47. Dahl, A.L. and R. Larsen. *Learning Dictionaries of Discriminative Image Patches*. in *BMVC*. 2011. Scotland, UK.
48. Huang, B., Z. Chen, P.-M. Wu, et al., *Fully automated delineation of gross tumor volume for head and neck cancer on PET-CT using deep learning: a dual-center study.* Contrast media & molecular imaging, 2018. **2018**.
49. Smith, R.L., S.J. Paisey, N. Evans, et al., *Deep learning pre-clinical medical image segmentation for automated organ-wise delineation of PET*, in *Annual Congress of the European Association of Nuclear Medicine*. 2018: Barcelona, Spain.2018.Barcelona, Spain.
50. Kostyszyn, D., T. Fechter, N. Bartl, et al., *Intraprostatic Tumour Segmentation on PSMA-PET Images in Patients with Primary Prostate Cancer with a Convolutional Neural Network.* Journal of Nuclear Medicine, 2020: p. jnumed. 120.254623.
51. Zhao, Y., A. Gafita, B. Vollnberg, et al., *Deep neural network for automatic characterization of lesions on 68 Ga-PSMA-11 PET/CT.* European journal of nuclear medicine and molecular imaging, 2020. **47**(3): p. 603-613.
52. Pantel, A.R. and D.A. Mankoff, *Molecular imaging to guide systemic cancer therapy: Illustrative examples of PET imaging cancer biomarkers.* Cancer letters, 2017. **387**: p. 25-31.
53. Ronneberger, O., P. Fischer, and T. Brox. *U-net: Convolutional networks for biomedical image segmentation*. in *International Conference on Medical image computing and computer-assisted intervention*. 2015. Munich, Germany: Springer.
54. Çiçek, Ö., A. Abdulkadir, S.S. Lienkamp, et al. *3D U-Net: learning dense volumetric segmentation from sparse annotation*. in *International conference on medical image computing and computer-assisted intervention*. 2016. Athens, Greece: Springer.
55. Milletari, F., N. Navab, and S.-A. Ahmadi. *V-net: Fully convolutional neural networks for volumetric medical image segmentation*. in *2016 Fourth International Conference on 3D Vision (3DV)*. 2016. Stanford, CA, USA: IEEE.
56. Zhou, Z., V. Sodha, M.M.R. Siddiquee, et al. *Models genesis: Generic autodidactic models for 3d medical image analysis*. in *International Conference on Medical Image Computing and Computer-Assisted Intervention*. 2019. Shenzhen, China: Springer.
57. Li, X., H. Chen, X. Qi, et al., *H-DenseUNet: hybrid densely connected UNet for liver and tumor segmentation from CT volumes.* IEEE transactions on medical imaging, 2018. **37**(12): p. 2663-2674.
58. Alom, M.Z., M. Hasan, C. Yakopcic, et al., *Recurrent residual convolutional neural network based on u-net (r2u-net) for medical image segmentation.* arXiv preprint arXiv:1802.06955, 2018.
59. Goodfellow, I., J. Pouget-Abadie, M. Mirza, et al. *Generative adversarial nets*. in *Advances in neural information processing systems*. 2014.

60. Roy, A.G., N. Navab, and C. Wachinger. *Concurrent spatial and channel 'squeeze & excitation' in fully convolutional networks*. in *International conference on medical image computing and computer-assisted intervention*. 2018. Springer.
61. Kamnitsas, K., C. Ledig, V.F. Newcombe, et al., *Efficient multi-scale 3D CNN with fully connected CRF for accurate brain lesion segmentation.* Medical image analysis, 2017. **36**: p. 61-78.
62. Oktay, O., J. Schlemper, L.L. Folgoc, et al., *Attention u-net: Learning where to look for the pancreas.* arXiv preprint arXiv:1804.03999, 2018.
63. Teramoto, A., H. Fujita, O. Yamamuro, et al., *Automated detection of pulmonary nodules in PET/CT images: Ensemble false-positive reduction using a convolutional neural network technique.* Medical physics, 2016. **43**(6Part1): p. 2821-2827.
64. Bi, L., J. Kim, A. Kumar, et al., *Automatic detection and classification of regions of FDG uptake in whole-body PET-CT lymphoma studies.* Computerized Medical Imaging and Graphics, 2017. **60**: p. 3-10.
65. van Tulder, G. and M. de Bruijne, *Representation learning for cross-modality classification*, in *Medical Computer Vision and Bayesian and Graphical Models for Biomedical Imaging*. 2016, Springer: Athens, Greece. p. 126-136.2016.Athens, Greece.
66. Yuan, C., M. Zhang, X. Huang, et al., *Diffuse Large B-cell Lymphoma Segmentation in PET-CT Images via Hybrid Learning for Feature Fusion.* Medical Physics, 2021.
67. Liu, Z., J. Mhlanga, R. Laforest, et al., *A Bayesian approach to tissue-fraction estimation for oncological PET segmentation.* Physics in Medicine & Biology, 2021(Special Issue on Early Career Researchers).
68. Lian, C., S. Ruan, T. Denoeux, et al. *Accurate tumor segmentation in FDG-PET images with guidance of complementary CT images*. in *2017 IEEE International Conference on Image Processing (ICIP)*. 2017. Beijing, China: IEEE.
69. Lian, C., H. Li, P. Vera, et al. *Unsupervised co-segmentation of tumor in PET-CT images using belief functions based fusion*. in *2018 IEEE 15th International Symposium on Biomedical Imaging (ISBI 2018)*. 2018. Washington, DC, USA: IEEE.
70. Sun, C., A. Shrivastava, S. Singh, et al. *Revisiting unreasonable effectiveness of data in deep learning era*. in *Proceedings of the IEEE international conference on computer vision*. 2017. Venice, Italy.
71. Willemink, M.J., W.A. Koszek, C. Hardell, et al., *Preparing medical imaging data for machine learning.* Radiology, 2020. **295**(1): p. 4-15.
72. Afshari, S., A. BenTaieb, Z. Mirikharaji, et al. *Weakly supervised fully convolutional network for PET lesion segmentation*. in *Medical Imaging 2019: Image Processing*. 2019. International Society for Optics and Photonics.
73. Hu, Y., M. Modat, E. Gibson, et al., *Weakly-supervised convolutional neural networks for multimodal image registration.* Medical image analysis, 2018. **49**: p. 1-13.
74. Zhou, Y., Y. Wang, P. Tang, et al. *Semi-supervised 3D abdominal multi-organ segmentation via deep multi-planar co-training*. in *2019 IEEE Winter Conference on Applications of Computer Vision (WACV)*. 2019. Waikoloa, HI, USA: IEEE.
75. Cheplygina, V., M. de Bruijne, and J.P. Pluim, *Not-so-supervised: a survey of semi-supervised, multi-instance, and transfer learning in medical image analysis.* Medical image analysis, 2019. **54**: p. 280-296.
76. Kamnitsas, K., C. Baumgartner, C. Ledig, et al. *Unsupervised domain adaptation in brain lesion segmentation with adversarial networks*. in *International conference on information processing in medical imaging*. 2017. Boone, NC, USA: Springer.
77. Lian, C., S. Ruan, T. Denœux, et al., *Spatial evidential clustering with adaptive distance metric for tumor segmentation in FDG-PET images.* IEEE Transactions on Biomedical Engineering, 2017. **65**(1): p. 21-30.
78. Lian, C., S. Ruan, T. Denœux, et al., *Joint tumor segmentation in PET-CT images using co-clustering and fusion based on belief functions.* IEEE Transactions on Image Processing, 2018. **28**(2): p. 755-766.
79. Hu, H., P. Decazes, P. Vera, et al., *Detection and segmentation of lymphomas in 3D PET images via clustering with entropy-based optimization strategy.* International journal of computer assisted radiology and surgery, 2019. **14**(10): p. 1715-1724.

80. Baur, C., B. Wiestler, S. Albarqouni, et al. *Deep autoencoding models for unsupervised anomaly segmentation in brain MR images*. in *International MICCAI Brainlesion Workshop*. 2018. Granada, Spain: Springer.
81. Wu, X., L. Bi, M. Fulham, et al. *Unsupervised Positron Emission Tomography Tumor Segmentation via GAN based Adversarial Auto-Encoder*. in *2020 16th International Conference on Control, Automation, Robotics and Vision (ICARCV)*. 2020. Shenzhen, China: IEEE.
82. Klyuzhin, I., Y. Xu, S. Harsini, et al., *Unsupervised background removal by dual-modality PET/CT guidance: application to PSMA imaging of metastases*, in *2021 SNMMI Annual meeting*. 2021: Washington DC.2021.Washington DC.
83. Soret, M., S.L. Bacharach, and I. Buvat, *Partial-volume effect in PET tumor imaging.* Journal of nuclear medicine, 2007. **48**(6): p. 932-945.
84. Rousset, O., A. Rahmim, A. Alavi, et al., *Partial volume correction strategies in PET.* PET clinics, 2007. **2**(2): p. 235-249.
85. Liu, Z., H.S. Moon, R. Laforest, et al., *Fully automated 3D segmentation of dopamine transporter SPECT images using an estimation-based approach.* arXiv preprint arXiv:2101.06729, 2021.
86. Sadik, M., E. Lind, E. Polymeri, et al., *Automated quantification of reference levels in liver and mediastinal blood pool for the Deauville therapy response classification using FDG-PET/CT in Hodgkin and non-Hodgkin lymphomas.* Clinical physiology and functional imaging, 2019. **39**(1): p. 78-84.
87. Wang, H., N. Zhang, L. Huo, et al., *Dual-modality multi-atlas segmentation of torso organs from [18 F] FDG-PET/CT images.* International journal of computer assisted radiology and surgery, 2019. **14**(3): p. 473-482.
88. Rister, B., D. Yi, K. Shivakumar, et al., *CT organ segmentation using GPU data augmentation, unsupervised labels and IOU loss.* arXiv preprint arXiv:1811.11226, 2018.
89. Rister, B., D. Yi, K. Shivakumar, et al., *CT-ORG, a new dataset for multiple organ segmentation in computed tomography.* Scientific Data, 2020. **7**(1): p. 1-9.
90. Yu, Y., P. Decazes, I. Gardin, et al., *3D lymphoma segmentation in PET/CT images based on fully connected CRFs*, in *Molecular Imaging, Reconstruction and Analysis of Moving Body Organs, and Stroke Imaging and Treatment*. 2017, Springer. p. 3-12.
91. Klyuzhin, I., G. Chausse, I. Bloise, et al., *Automated deep segmentation of healthy organs in PSMA PET/CT images*, in *2021 SNMMI Annual meeting*. 2021: Washington DC.2021.Washington DC.
92. Zhang, P., Y. Zhong, Y. Deng, et al., *A survey on deep learning of small sample in biomedical image analysis.* arXiv preprint arXiv:1908.00473, 2019.
93. Isola, P., J.-Y. Zhu, T. Zhou, et al. *Image-to-image translation with conditional adversarial networks*. in *Proceedings of the IEEE conference on computer vision and pattern recognition*. 2017. Honolulu, HI, USA.
94. Bi, L., J. Kim, A. Kumar, et al., *Synthesis of positron emission tomography (PET) images via multi-channel generative adversarial networks (GANs)*, in *molecular imaging, reconstruction and analysis of moving body organs, and stroke imaging and treatment*. 2017, Springer. p. 43-51.
95. Ben-Cohen, A., E. Klang, S.P. Raskin, et al. *Virtual PET images from CT data using deep convolutional networks: initial results*. in *International workshop on simulation and synthesis in medical imaging*. 2017. Québec City, QC, Canada: Springer.
96. Ben-Cohen, A., E. Klang, S.P. Raskin, et al., *Cross-modality synthesis from CT to PET using FCN and GAN networks for improved automated lesion detection.* Engineering Applications of Artificial Intelligence, 2019. **78**: p. 186-194.
97. Kazeminia, S., C. Baur, A. Kuijper, et al., *GANs for medical image analysis.* Artificial Intelligence in Medicine, 2020. **109**: p. 101938.
98. Segars, W.P., G. Sturgeon, S. Mendonca, et al., *4D XCAT phantom for multimodality imaging research.* Medical physics, 2010. **37**(9): p. 4902-4915.
99. Leung, K., W. Marashdeh, R. Wray, et al., *A deep-learning-based fully automated segmentation approach to delineate tumors in FDG-PET images of patients with lung cancer.* Journal of Nuclear Medicine, 2018. **59**(supplement 1): p. 323-323.
100. Leung, K.H., W. Marashdeh, R. Wray, et al., *A physics-guided modular deep-learning based automated framework for tumor segmentation in PET.* Physics in Medicine & Biology, 2020. **65**(24): p. 245032.
101. Liu, Z., R. Laforest, J. Mhlanga, et al. *Observer study-based evaluation of a stochastic and physics-based method to generate oncological PET images*. in *Medical Imaging 2021: Image Perception,*

*Observer Performance, and Technology Assessment*. 2021. International Society for Optics and Photonics.

102. Andrearczyk, V., V. Oreiller, M. Jreige, et al. *Overview of the HECKTOR challenge at MICCAI 2020: automatic head and neck tumor segmentation in PET/CT*. in *3D Head and Neck Tumor Segmentation in PET/CT Challenge*. 2020. Springer.
103. Vallieres, M., E. Kay-Rivest, L.J. Perrin, et al., *Radiomics strategies for risk assessment of tumour failure in head-and-neck cancer.* Scientific reports, 2017. **7**(1): p. 1-14.
104. Kadota, K., J.-i. Nitadori, C.S. Sima, et al., *Tumor spread through air spaces is an important pattern of invasion and impacts the frequency and location of recurrences after limited resection for small stage I lung adenocarcinomas.* Journal of Thoracic Oncology, 2015. **10**(5): p. 806-814.
105. Dou, T.H., T.P. Coroller, J.J. van Griethuysen, et al., *Peritumoral radiomics features predict distant metastasis in locally advanced NSCLC.* PloS one, 2018. **13**(11): p. e0206108.
106. Hatt, M., C.C. Le Rest, A. Turzo, et al., *A fuzzy locally adaptive Bayesian segmentation approach for volume determination in PET.* IEEE transactions on medical imaging, 2009. **28**(6): p. 881-893.
107. Mehta, P., V. Sandfort, D. Gheysens, et al. *Segmenting The Kidney On CT Scans Via Crowdsourcing*. in *2019 IEEE 16th International Symposium on Biomedical Imaging (ISBI 2019)*. 2019. Venice, Italy: IEEE.
108. Heim, E., T. Roß, A. Seitel, et al., *Large-scale medical image annotation with crowd-powered algorithms.* Journal of Medical Imaging, 2018. **5**(3): p. 034002.
109. Surowiecki, J., *The wisdom of crowds*. 2005, NewYork: Doubleday; Anchor.NewYork.
110. Taha, A.A. and A. Hanbury, *Metrics for evaluating 3D medical image segmentation: analysis, selection, and tool.* BMC medical imaging, 2015. **15**(1): p. 1-28.
111. Renard, F., S. Guedria, N. De Palma, et al., *Variability and reproducibility in deep learning for medical image segmentation.* Scientific Reports, 2020. **10**(1): p. 1-16.
112. Jha, A.K., B. Caffo, and E.C. Frey, *A no-gold-standard technique for objective assessment of quantitative nuclear-medicine imaging methods.* Physics in Medicine & Biology, 2016. **61**(7): p. 2780.
113. Sander, J., B.D. de Vos, J.M. Wolterink, et al. *Towards increased trustworthiness of deep learning segmentation methods on cardiac MRI*. in *Medical Imaging 2019: Image Processing*. 2019. International Society for Optics and Photonics.
114. Zhu, Y., F. Yousefirizi, Z. Liu, et al., *Comparing clinical evaluation of PET segmentation methods with reference-based metrics and no-gold-standard evaluation technique*, in *SNMMI 2021*. 2021, Soc Nuclear Med: Washington DC.2021.Washington DC.
115. Jha, A.K., M.A. Kupinski, J.J. Rodríguez, et al. *Evaluating segmentation algorithms for diffusion-weighted MR images: a task-based approach*. in *Medical Imaging 2010: Image Perception, Observer Performance, and Technology Assessment*. 2010. International Society for Optics and Photonics.
116. Jha, A.K., M.A. Kupinski, J.J. Rodriguez, et al., *Task-based evaluation of segmentation algorithms for diffusion-weighted MRI without using a gold standard.* Physics in Medicine & Biology, 2012. **57**(13): p. 4425.
117. Lebenberg, J., I. Buvat, A. Lalande, et al., *Nonsupervised ranking of different segmentation approaches: application to the estimation of the left ventricular ejection fraction from cardiac cine MRI sequences.* IEEE Transactions on Medical Imaging, 2012. **31**(8): p. 1651-1660.
118. Jha, A.K., N. Song, B. Caffo, et al. *Objective evaluation of reconstruction methods for quantitative SPECT imaging in the absence of ground truth*. in *Medical Imaging 2015: Image Perception, Observer Performance, and Technology Assessment*. 2015. International Society for Optics and Photonics.
119. Jha, A.K., E. Mena, B.S. Caffo, et al., *Practical no-gold-standard evaluation framework for quantitative imaging methods: application to lesion segmentation in positron emission tomography.* Journal of Medical Imaging, 2017. **4**(1): p. 011011.
120. Liu, J., Z. Liu, H.S. Moon, et al., *A no-gold-standard technique for objective evaluation of quantitative nuclear-medicine imaging methods in the presence of correlated noise.* Journal of Nuclear Medicine, 2020. **61**(supplement 1): p. 523-523.
121. Zhu, Y., Z. Liu, M. Bilgel, et al., *No-gold-standard evaluation of partial volume compensation methods for brain PET*, in *SNMMI 2021*. 2021, Soc Nuclear Med: Washington DC.2021.Washington DC.

122. Jha, A. and E. Frey. *Incorporating prior information in a no-gold-standard technique to assess quantitative SPECT reconstruction methods*. in *International Meeting on Fully 3D reconstruction in Radiology and Nuclear Medicine*. 2015. Newport, Rhode Island, USA.
123. Buvat, I. and F. Orlhac, *The TRUE checklist for identifying impactful AI-based findings in nuclear medicine: is it True? Is it Reproducible? Is it Useful? Is it Explainable?* Journal of Nuclear Medicine, 2021. **Vol. 62**(7).
124. Chang, K., N. Balachandar, C. Lam, et al., *Distributed deep learning networks among institutions for medical imaging.* Journal of the American Medical Informatics Association, 2018. **25**(8): p. 945-954.
125. National Academies of Sciences, E. and Medicine, *Reproducibility and replicability in science*. 2019, Washington, DC: National Academies Press.Washington, DC.
126. Bi, W.L., A. Hosny, M.B. Schabath, et al., *Artificial intelligence in cancer imaging: clinical challenges and applications.* CA: a cancer journal for clinicians, 2019. **69**(2): p. 127-157.
127. Baumgartner, R., A. Joshi, D. Feng, et al., *Statistical evaluation of test-retest studies in PET brain imaging.* EJNMMI research, 2018. **8**(1): p. 1-9.
128. Bland, J.M. and D.G. Altman, *Measuring agreement in method comparison studies.* Statistical methods in medical research, 1999. **8**(2): p. 135-160.
129. Lodge, M.A., *Repeatability of SUV in oncologic 18F-FDG PET.* Journal of Nuclear Medicine, 2017. **58**(4): p. 523-532.
130. Mongan, J., L. Moy, and J. Charles E. Kahn, *Checklist for Artificial Intelligence in Medical Imaging (CLAIM): A Guide for Authors and Reviewers.* Radiology: Artificial Intelligence, 2020. **2**(2): p. e200029.
131. Kendall, A. and Y. Gal, *What uncertainties do we need in bayesian deep learning for computer vision?* arXiv preprint arXiv:1703.04977, 2017.
132. Wang, G., W. Li, M. Aertsen, et al., *Aleatoric uncertainty estimation with test-time augmentation for medical image segmentation with convolutional neural networks.* Neurocomputing, 2019. **338**: p. 34-45.
133. Lakshminarayanan, B., A. Pritzel, and C. Blundell, *Simple and Scalable Predictive Uncertainty Estimation using Deep Ensembles.* Advances in Neural Information Processing Systems, 2017. **30**.
134. Kwon, Y., J.-H. Won, B.J. Kim, et al., *Uncertainty quantification using bayesian neural networks in classification: Application to ischemic stroke lesion segmentation.* Computational Statistics & Data Analysis, 2018. **142**.
135. Stephens, K., *FDA Releases Artificial Intelligence/Machine Learning Action Plan.* AXIS Imaging News, 2021.
136. Dean, J., G.S. Corrado, R. Monga, et al., *Large scale distributed deep networks*, in *Proceedings of NIPS*. 2012. p. 1232–1240.2012.
137. Su, H. and H. Chen, *Experiments on parallel training of deep neural network using model averaging.* arXiv preprint arXiv:1507.01239, 2015.
138. Kairouz, P., H.B. McMahan, B. Avent, et al., *Advances and open problems in federated learning.* arXiv preprint arXiv:1912.04977, 2019.
139. Zerka, F., S. Barakat, S. Walsh, et al., *Systematic review of privacy-preserving distributed machine learning from federated databases in health care.* JCO clinical cancer informatics, 2020. **4**: p. 184-200.
140. Iglesias, J.E., C.-Y. Liu, P. Thompson, et al. *Agreement-based semi-supervised learning for skull stripping*. in *International Conference on Medical Image Computing and Computer-Assisted Intervention*. 2010. Beijing, China: Springer.
141. Li, X., L. Yu, H. Chen, et al., *Semi-supervised skin lesion segmentation via transformation consistent self-ensembling model.* arXiv preprint arXiv:1808.03887, 2018.
142. Kervadec, H., J. Dolz, É. Granger, et al. *Curriculum semi-supervised segmentation*. in *International Conference on Medical Image Computing and Computer-Assisted Intervention*. 2019. Shenzhen, China: Springer.
143. Manhaeve, R., S. Dumancic, A. Kimmig, et al., *Deepproblog: Neural probabilistic logic programming.* Advances in Neural Information Processing Systems, 2018. **31**: p. 3749-3759.
144. Došilović, F.K., M. Brčić, and N. Hlupić. *Explainable artificial intelligence: A survey*. in *2018 41st International convention on information and communication technology, electronics and microelectronics (MIPRO)*. 2018. Opatija, Croatia: IEEE.
145. Folio, L.R., L.B. Machado, and A.J. Dwyer, *Multimedia-enhanced radiology reports: concept, components, and challenges.* RadioGraphics, 2018. **38**(2): p. 462-482.